\documentclass[12pt,preprint]{aastex}

\slugcomment{To appear in AJ }

\shorttitle{Recent star formation in clusters of galaxies}
\shortauthors{Reverte et al.}

\begin{document}


\title{Recent star formation in clusters of galaxies: extreme compact starbursts in A539 and A634}



\author{D. Reverte, J. M. V\'{\i}lchez, J.D. Hern\'{a}ndez-Fern\'{a}ndez and
  J. Iglesias-P\'{a}ramo} 
\affil{Instituto de Astrof\'{\i}sica de Andaluc\'\i a-CSIC, Apdo. 3004,
    Granada, 18080 Spain}




\begin{abstract}
We report on the detection of two H$\alpha$-emitting extreme compact
objects from deep images of the Abell~634 and Abell~539 clusters
of galaxies at $z \sim 0.03$.
Follow up long slit spectroscopy of these two unresolved sources revealed
that they are members of their respective clusters showing H{\sc ii} type
spectra.  The luminosity and the extreme equivalent width of H$\alpha
+$[N{\sc ii}]  
measured for these sources, together with their very compact appearance,
has raised the question about the origin of these intense starbursts in the
cluster 
environment. We propose the compact starburst in Abell~539 resulted from
the compression of  the interstellar gas of a dwarf galaxy  when entering the
cluster core; while the starburst galaxy in Abell~634 is likely to be the
result of a 
galaxy-galaxy  interaction, illustrating the preprocessing of galaxies during
their infall towards the central regions of clusters.
The contribution of these compact star-forming dwarf galaxies to the star
formation history of galaxy clusters is discussed, as well as  a possible
link with the recently discovered early-type ultra-compact dwarf galaxies.
We note that these extreme objects will be rarely
detected in normal magnitude-limited optical or NIR surveys, mainly due to
their low stellar masses (of the order of $10^{6}$~M$_{\odot}$), whereas
they will easily  show up in dedicated H$\alpha$ surveys given the high
equivalent width of their emission lines.

\end{abstract}



\keywords{galaxies: clusters of galaxies, ICM ---
galaxies: starburst, abundancies, star-formation}


\section{Introduction}

The understanding of the role played by the environment on galaxy evolution
still remains a major issue. A key aspect to understanding the evolution of
galaxies in 
clusters is the knowledge of  their star formation (SF) history, intimately
related to their morphology and gas content. Indeed,  it has been
reported that the global star formation rates of spiral galaxies located in
the innermost regions of nearby and intermediate redshift clusters appear
strongly depressed as compared to the results found for similar galaxies at
larger clustercentric radii (e.g. \citealt{bal98,bal04,gav02,lew02}).
Less information is available in the
literature with respect to the evolution of the SF activity of the population
of dwarf galaxies in clusters. To date, only a handful of works have
dealt with the study of the SF activity in the population of star-forming
dwarf  and irregular galaxies in nearby clusters
(e.g. \citealt{gal89,drnk91,vil95,duc01,bos02b,ige03,lee03,vil03}).

Several physical mechanisms have been invoked to explain the influence of the
cluster environment on the evolution of star-forming galaxies  (see
\citealt{bos06}). Gas-rich late-type galaxies falling for the first time into
the intra-cluster medium (ICM) of a rich cluster can  suffer compression of
their interstellar medium by ram pressure, triggering star formation bursts;
this process can be followed by the stripping of  their external gaseous
component, thus inducing a quenching of their star formation activity. In
addition, gravitational interactions could  give rise to different kinds  of
tidal interactions: with other galaxies, with the cluster potential,
``harassment'' \citep{moo00}. Other important processes include
``starvation'' of galactic gas component and ``preprocessing'' of galaxies
in falling groups into clusters \citep{pog04}. Overall, SF activity is
expected to be more efficient in high velocity objects at the periphery of
the clusters, as shown by numerical simulations \citep{fuj99, mor00}. From
the observational point of view, it is not yet clear if the SF activity of
cluster dwarf irregulars may vary along the clustercentric radius. It seems
clear that dwarf galaxies and large spirals should show different responses
to the action of cluster tidal fields as a physical consequence of their
different mass concentrations \citep{moo99}. To date, there is not
enough observational information available on the evolution of star-forming
dwarf galaxies in clusters, mostly a consequence of the magnitude-limited
searches.

Isolated intergalactic H{\sc ii} regions have been recently found in the
vicinity of cluster spirals, thus providing evidence for their origin in
tidal interactions and from previously processed material
\citep{ger02,cor04}. Furthermore, a population of intergalactic planetary
nebulae 
(PN) has been reported to exist in Virgo \citep{fre00,arn03} and in Coma
\citep{ger02,ger05} clusters. Obviously
the luminosity contrast of the emission lines of H{\sc ii} regions and
PN and their relative compactness has allowed them to be detected in nearby
clusters, but not in more distant ones given their low luminosities.

A new class of ultra compact dwarf galaxies (UCDGs) in the nearby clusters
Fornax \citep{drnk00} and Virgo \citep{jon06} has been recently
discovered. These 
galaxies show extremely compact surface brightness profiles and sizes
slightly larger than those of stellar clusters, a few tens of pc, with
  absolute B magnitude  -13 $\lesssim M_{B} \lesssim$ -11  \citep{drnk04},
  and typically also present spectra lacking emission-lines. Two explanations
for the nature of these galaxies have been proposed: either they are the
successors of tidally stripped dEs, or they have originated from merged
young massive clusters of tidal origin \citep{mie06}. Up to now, the typical
UCDGs reported in clusters do not appear to show recent star formation
activity,  thus the possibility of analogous star-forming UCDGs in the
cluster environment could shed light on the nature and evolution of dwarf
galaxies. 

A rather unexplored possibility is the search for very compact,
actively star-forming dwarf galaxies in clusters. These galaxies should show
emission lines with high equivalent widths, as shown by PN and  H{\sc ii}
regions, though their expected intrinsic luminosity would be much larger than
in the cases of PN and H{\sc ii} regions, thus favouring their detection in
clusters. Besides the work on Virgo \citep{ger02} and Fornax
\citep{drnk01}, there is recent evidence for ``missing'' compact galaxies
in the  local field, i.e. galaxies that have been misclassified as stars due
to their compactness by standard star-galaxy separation techniques, in the
Millennium Galaxy Catalogue \citep{lis06}.

In this work we report on the detection and properties of two very compact
strong starbursts that have been found associated to the clusters of galaxies
Abell~539 and Abell~634. 
 These two objects were discovered by visual inspection of some H$\alpha$
  frames covering approximately 1$^{o} \times$ 1$^{o}$ on the basis
  of their compactness and strong H$\alpha$ emission.
The parent clusters are approximately at the same
distance ($z \approx 0.03$) and show rather different properties: whereas
Abell~539 is a rich X-ray luminous cluster, Abell~634  is a
poor, disperse cluster, for which only an upper limit in X-ray  by
ROSAT is available, probably indicating an ongoing galaxy assembling
from the cluster outskirts. 
The adopted heliocentric velocities (velocity dispersions) for A539 and 
A634 are 8514~km~s$^{-1}$ ($\sigma_{A539} = 629$~km~s$^{-1}$) and 
7945~km~s$^{-1}$ ($\sigma_{A634} = 391$~km~s$^{-1}$) respectively
\citep{str99}. 
 Thus assuming
standard cosmology with $H_{0} = 70$~km~s$^{-1}$~Mpc$^{-1}$,  $\Omega_{M} =
0.3$ and $\Omega_{VAC} = 0.7$, the distances to the clusters adopted
throughout this paper are 124.2 (Abell~539) and 115.7~Mpc (Abell~634). 

This paper is organized as follows: Section~2 describes the observations and
data handling. Section~3 enumerates the main observational properties of the
two galaxies and in section~4 we  discuss their main properties, origin and
evolutionary stage. Finally, in Section~5 we present  our conclusions and
final remarks about the relative importance of these kinds of objects for
the SFR budget of nearby clusters of galaxies.

\section{Observations}

\subsection{Imaging}

H$\alpha +$continuum imaging of the two starburst  sources DRP-A539a and
DRP-A634a, were obtained with the Wide Field Camera (WFC) attached to the
Prime Focus of the 2.5m Isaac Newton Telescope (INT), at the Observatorio de
El Roque de los Muchachos (ORM), La Palma, Spain, in December 2002 and 2003,
and 
February 2003.  The WFC consists of an array of four thinned AR coated
EEV $\rm 4k \times 2k$ devices, plus a fifth used for autoguiding. The pixel
scale is 0.33 arcsec pixel$^{-1}$, which gives a total field of view of about
$34' \times 34'$. Given the  particular arrangement of the detectors, a
squared area of about $11' \times 11'$ is lost in the top right corner of the
field. Given the redshift of these objects, an ON-band narrow filter
($\lambda_{0} = 6725$\AA, $\Delta \lambda_{FWHM} = 85$\AA) to
isolate the H$\alpha$ emission and an OFF-band narrow filter ($\lambda_{0} =
6563$\AA, $\Delta \lambda_{FWHM} = 95$\AA) to measure the 
continuum emission corresponding to the redshifted H$\alpha$ line were
used. At least 
four different exposures, slightly dithered to remove cosmic rays, were
obtained for each position with each filter. The typical seeing of our frames
was between 0.8 and 1.5 arcsec except for a few nights of the 2002 run that
were around 2 arcsec. Table~\ref{image_log} shows a log of the imaging
observations. 

Data reduction was performed using the standard software package
IRAF\footnote{Image Reduction and Analysis Facility, written and
supported at the National Optical Astronomy Observatories.} following
the standard procedure of bias correction, flat-fielding and flux
calibration.
In order to properly subtract the continuum from the H$\alpha$ emission,
 the counts of the OFF-band frames were scaled so that the counts of
(non saturated) field stars were the same in ON-band and OFF-band frames.

ON-band observations of the spectrophotometric standard stars G191-B2B,
PG0934+554, PG0834+546, BD+33 2642, Feige 56 and Feige 67 were taken to
perform the flux calibration of our objects. The accuracy of the zero
point was 0.05 mag. 
The photometry of the ON-band and OFF-band sources was performed with
the IRAF task qphot, and consisted of circular aperture photometry until
convergence of the growth curve was achieved.

\begin{figure}
\centering
\includegraphics[width=10.1cm]{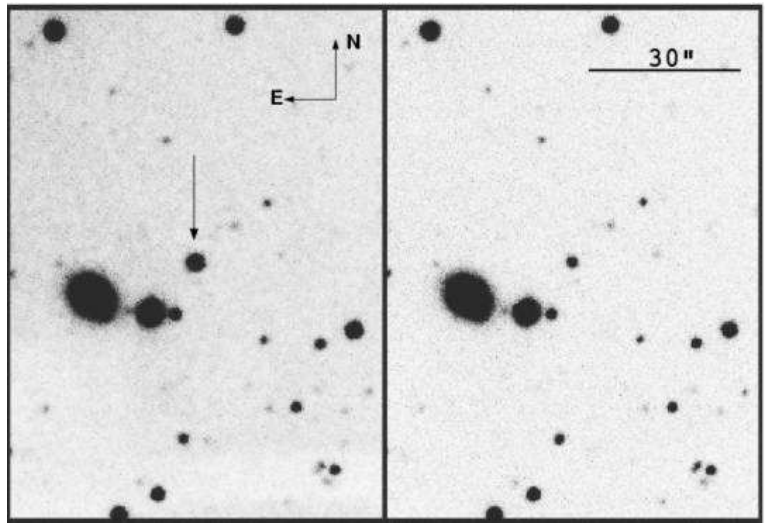}
\caption{ON-band (H$\alpha$ - left) and OFF-band (red continuum - right)
  images of DRP-A539a region. The arrow indicates the position of the
  object.} 
\label{a539ha}

\end{figure}

\begin{figure}
\centering
\includegraphics[width=14cm]{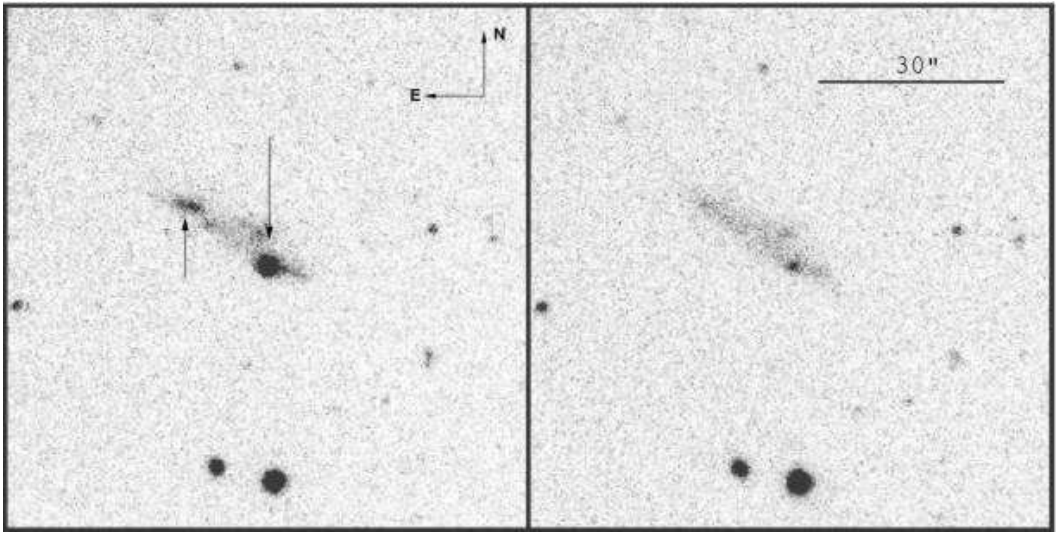}
\caption{ON-band (H$\alpha$ - left) and OFF-band (red continuum - right)
  images of DRP-A634a region. The downward arrow indicates the position of
  DRP-A634a. The upward arrow points to {LSB-A634a{\_}knot . The
    low surface brightness structure apparent in the right panel is
    LSB-A634a}.}   
\label{a634ha}
\end{figure}

Both objects, DRP-A539a and DRP-A634a, appear very compact and much
brighter in ON-band than in OFF-band, as shown in figs \ref{a539ha} and
\ref{a634ha}. While DRP-A539a 
appears to be an isolated source, 
 a diffuse low surface brightness structure has been detected extending
  North East from DRP-A634a. Hereafter,
  the compact source will be referred to as DRP-A634a, whereas the diffuse low surface
  brightness structure will be named LSB-A634a. 
Some knotty faint H$\alpha$ emission can be seen at the northern tip of LSB-A634a, which 
will be referred to as LSB-A634a{\_}knot.
Table~\ref{phot_pro} shows the relevant photometric properties of these objects, as
measured from our H$\alpha$ and continuum frames. The luminosities have been
corrected for Galactic extinction following \citet{sch98}, and the
\citet{car89} extinction curve. The H$\alpha +$[N{\sc ii}] luminosity of
LSB-A634a{\_}knot has been estimated assuming that it is located at 
the same distance as DRP-A634a\footnote{The possible origins  of these
  objects and their environment will be discussed in Section~3.}.  

\subsection{Spectroscopy}

Long slit spectroscopy of DRP-A539a, DRP-A634a and LSB-A634a was obtained with the
Alhambra Faint Object Spectrograph and Camera (ALFOSC) attached to the 2.5m
Nordic Optical Telescope (NOT) at the ORM on 19th November 2004 and 9th
December 2004. Grisms \#8 and 
\#14 were used, giving useful spectral ranges of $\lambda$5825--8350\AA\ and
$\lambda$3275--6125\AA\ respectively. 
 The spatial resolution across the slit was 0.19''~pix$^{-1}$.
During the first night, the  slit width was set to 1.2''
resulting in an effective spectral resolution of 7.1~\AA\ and 
a 1800 s exposure for each spectral range was
obtained for DRP-A634a. 
The second night was devoted to DRP-A539a and LSB-A634a. Due to an improvement in
the weather conditions the slit width was set to 0.4'', yielding a spectral
resolution of 2.8~\AA. A total of 
3$\times$1800~s exposures for each spectral range of DRP-A539a were taken with
the slit oriented along the parallactic angle. Finally, a 600 s exposure using
grism \#8 was performed in order to observe the extended emission of
LSB-A634a; in this case, the slit was centered on DRP-A634a and was carefully
oriented at an angle of 54$^{o}$ (from North to East). 
Due to an increase in the humidity, this exposure was stopped after 600~s.
Table~\ref{spectra_log} shows the log of spectroscopic observations.

Data reduction was performed using IRAF, following the standard
procedure of bias correction, flat-fielding and wavelength and flux
calibration. 
1-D spectra of DRP-A539a and DRP-A634a were extracted by adding the flux 
  in the spatial sections along the slit 
which maximize their signal to noise ratios. A total of 10 and 7 pixels were
added for DRP-A539a and DRP-A634a respectively.
The same procedure was followed to
  extract a 1-D spectrum for LSB-A634a. As a consequence of its low surface
  brightness only a faint emission line, spatially corresponding to
  LSB-A634a{\_}knot, was obtained after adding a total of 21 pixels. As
  indicated below, this emission line was identified as H$\alpha$.

\begin{figure}
\centering
\includegraphics[width=14cm]{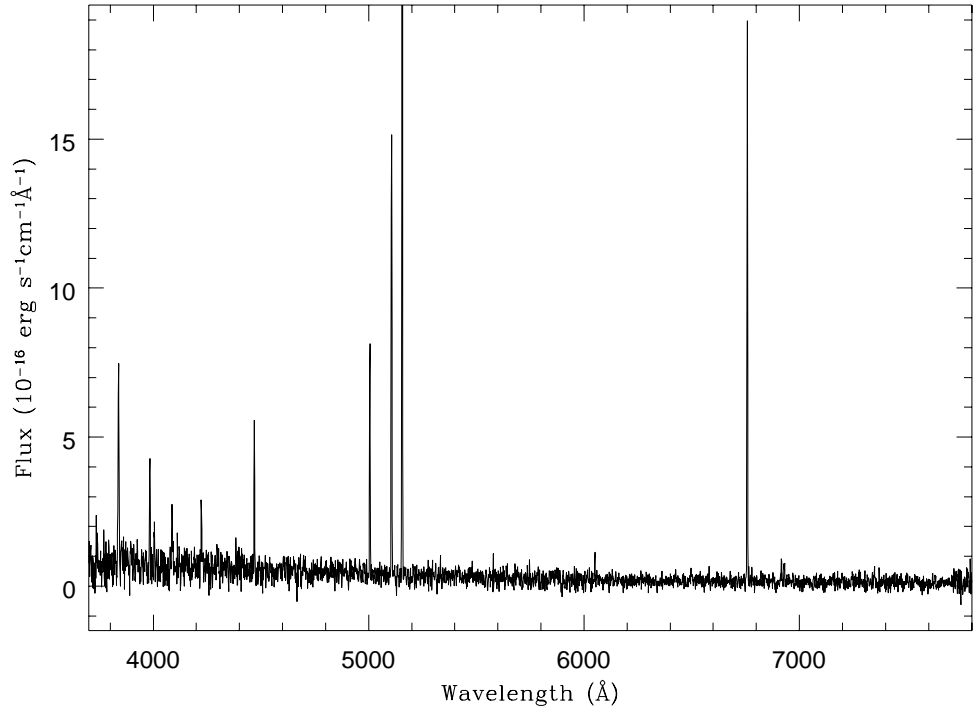}
\caption{Combined spectrum of DRP-A539a scaled to [O{\sc iii}]$\lambda$ 4959 {\AA}.
}
\label{a539sp}
\end{figure}

\begin{figure}
\centering
\includegraphics[width=14cm]{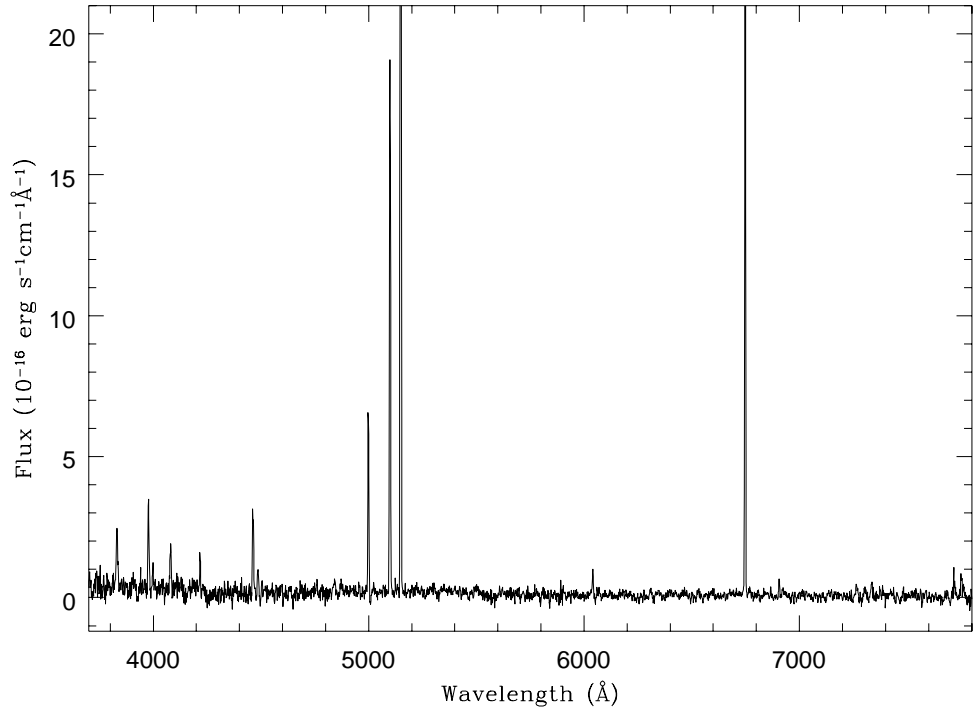}
\caption{Same as Figure~\ref{a539sp} for DRP-A634a.}
\label{a634sp}
\end{figure}

Both nights were only partially photometric, so an absolute spectrophotometric
flux calibration was not attempted. However, the spectrophotometric
standard star Hiltner~600 was observed before and after each object
with each grism, thus a relative calibration of the spectra in physical
units was performed. The spectra corresponding to the red grism were scaled
to the blue ones by using the continuum level and the flux of the
[He{\sc i}]$\lambda$5876\AA\ line, present in the blue and red spectra.
The scaling factors were found to be $\sim 1$ within an error bar
of $\sim 20$\%. Beyond 7600\AA\ fringing effects begin to be
noticeable and data at longer wavelengths is ineffective.
Figures~\ref{a539sp} and \ref{a634sp} show the combined
(red grism $+$ blue grism) spectra of DRP-A539a and DRP-A634a
respectively. Both spectra are dominated by narrow emission
lines and show a very faint underlying continuum.

The emission lines were measured with the IRAF task splot.
The errors of the line fluxes are estimated from the
standard deviation of a series of independent repeated measurements, 
sampling the adjacent continuum for each line. In order
to calculate the extinction, the Balmer decrement
was computed using the H$\alpha$, H$\beta$, H$\gamma$ and H$\delta$
line fluxes and  compared to their theoretical values
\citep{hum87}. Given the low continuum shown by both objects, no 
correction for underlying absorption was performed.  

Radial velocities were computed from a sigma-weighted average of the redshifts
corresponding to the individual emission lines. After applying the
heliocentric corrections, values of $v_{rad} = 8940 \pm 40$~km~s$^{-1}$ and
$8470 \pm 30$~km~s$^{-1}$ were found for DRP-A539a and
DRP-A634a respectively. The quoted errors correspond to the standard deviations of
the velocities derived from individual lines.
 The measured velocities are offset by about 500~km~s$^{-1}$ from the
  mean heliocentric velocities adopted for the parent clusters. As will be
  discussed in Sections~4.1 and 4.2, the projected positions of our two
  objects with respect to the center of the clusters, together with their
  redshifts, are consistent with their corresponding cluster memberships.

\begin{figure}
\centering
\includegraphics[width=11cm]{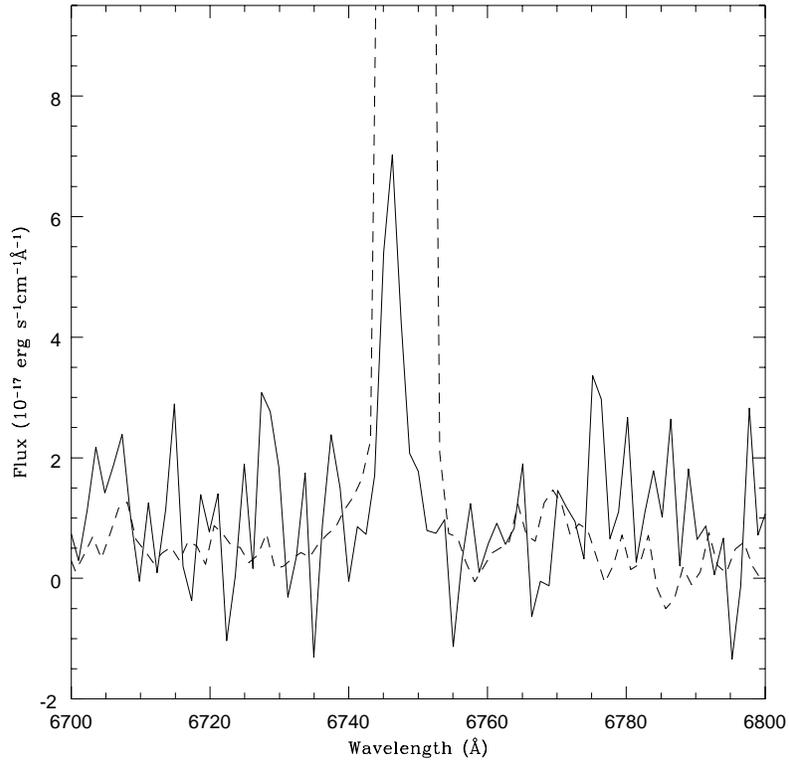}
\caption{ The solid line shows the spectrum of LSB-A634a{\_}knot,
  spatially binned to increase the signal to noise ratio. The dashed line
  shows the spectrum of DRP-A634a around the H$\alpha$ line.  The scale of
  the flux axis has been set to show the H$\alpha$ line of the knot component.}
\label{spec_lsb}
\end{figure}

 In the spectrum of LSB-A634a{\_}knot, shown in fig. \ref{spec_lsb},
  despite the low signal to noise ratio, an emission line was 
  detected at the 3.5 $\sigma$ level. This line, centered at $\lambda =
  6747$\AA\, is almost coincident with the wavelength of the H$\alpha$ line
  of DRP-A634a. Assuming that this line effectively corresponds to H$\alpha$,
  a radial velocity of 8390 km~s$^{-1}$ is inferred 
  for LSB-A634a{\_}knot, after correcting for heliocentric relative motions.

 In table~\ref{spec_pro} we present the spectroscopic properties of
DRP-A539a and DRP-A634a:
 reddening corrected line fluxes, reddening  coefficient $C(H\beta)$,
 equivalent widths of H$\beta$, H$\alpha$ and [O{\sc ii}], 
H$\beta$ flux, as well as the fluxes of the most prominent emission lines
relative to H$\beta$. As can be seen, the values reported for the
H$\alpha$ equivalent width, though slightly larger than the ones derived from
H$\alpha$ imaging are consistent within the errors.

Table \ref{spec_pro} also shows the electron temperatures and oxygen
abundances derived using {\sl temden} and {\sl ionic} tasks in the 
IRAF {\sl nebular} package in STSDAS. Oxygen abundances were derived
directly from spectral lines of [O{\sc ii}]$\lambda$3727, 
[O{\sc iii}]$\lambda\lambda$4959,5007 and using their electron 
temperatures from the measurements of [O{\sc iii}]$\lambda$4363.

\section{Results}

\begin{figure}
\centering
\includegraphics[width=11cm]{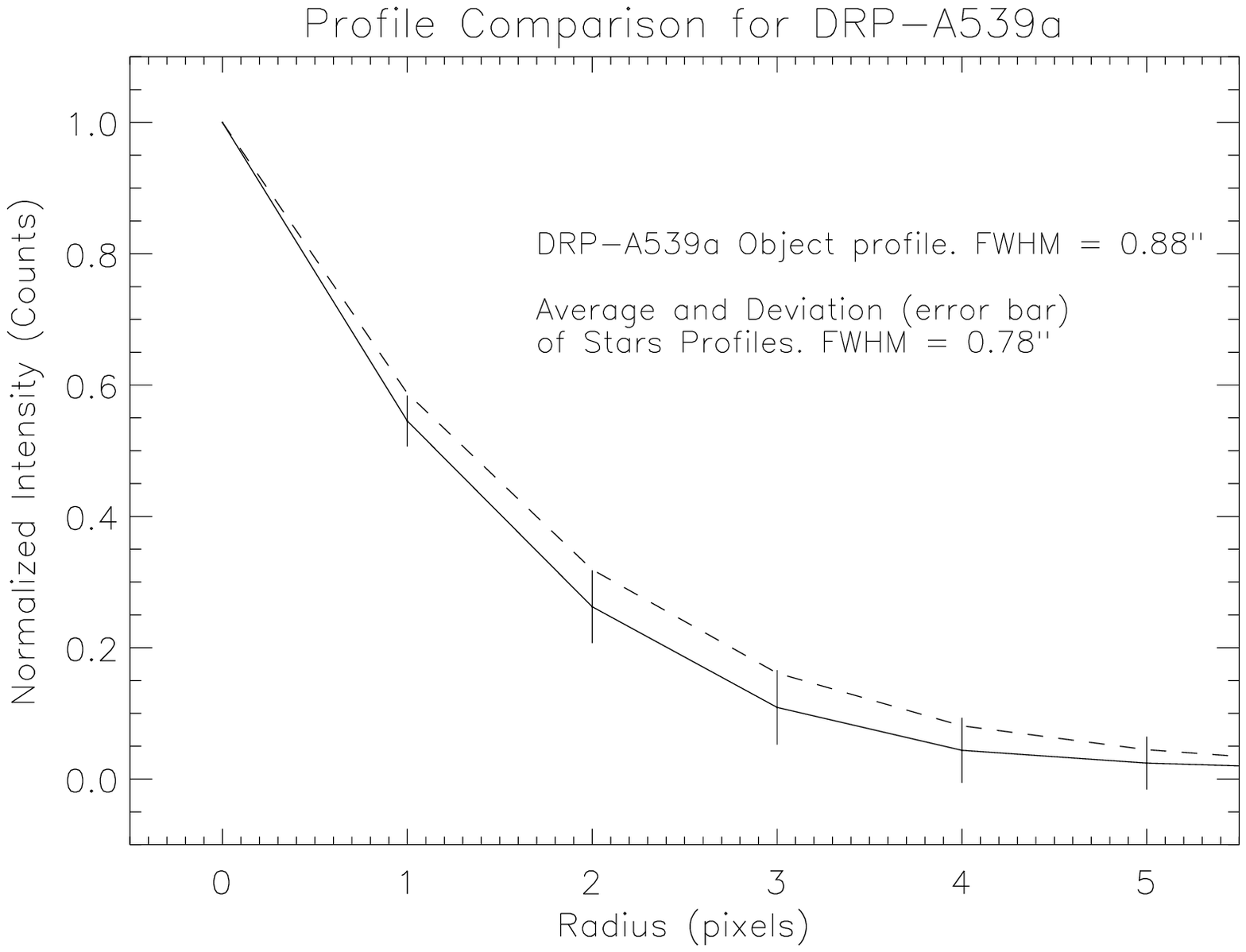}
\includegraphics[width=11cm]{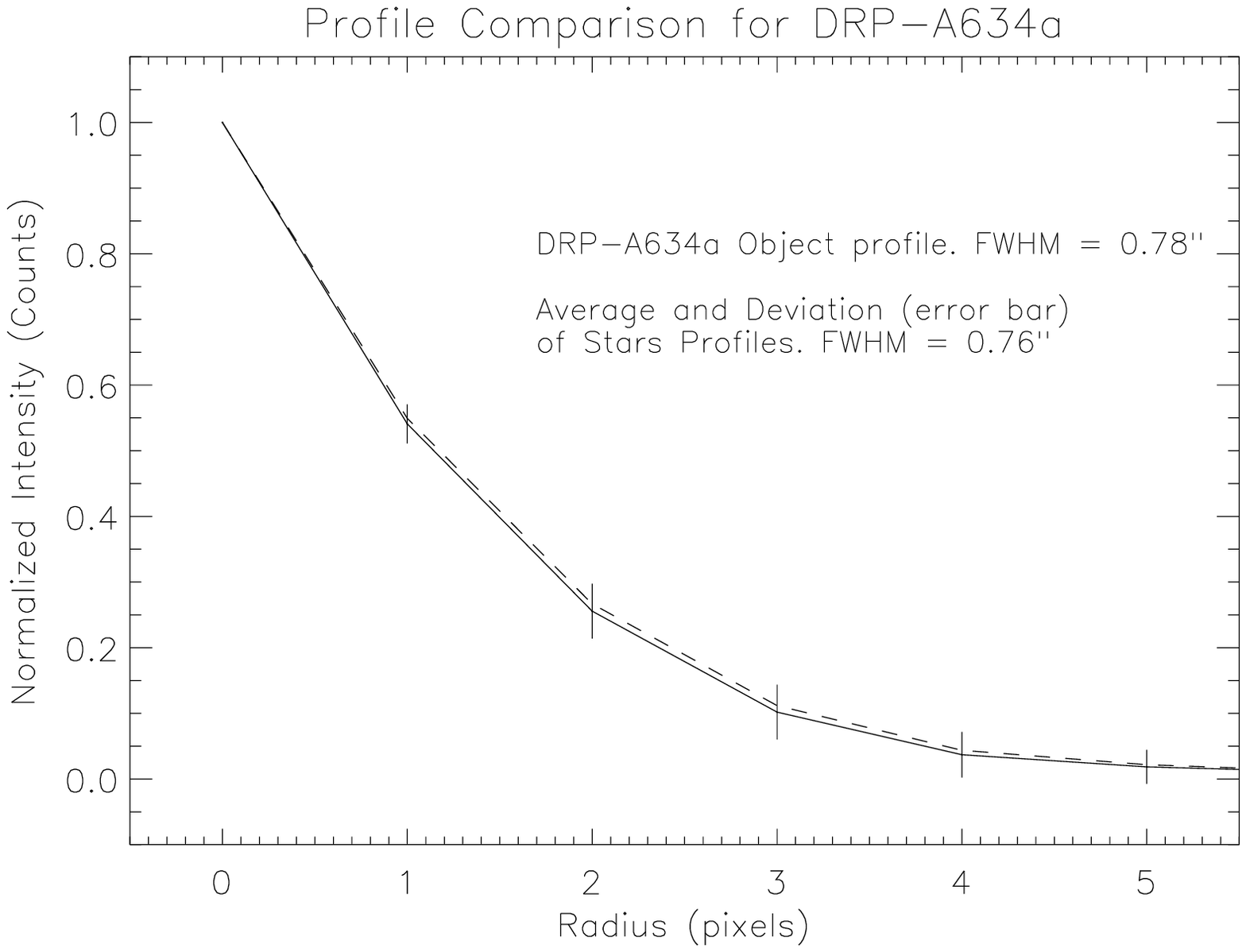}
\caption{Radial profiles of DRP-A539a (upper panel) and DRP-A634a (lower
  panel) indicated by the dashed lines. The solid lines correspond
to the stellar profiles averaged over a large number of stars for each frame.
}
\label{profiles}
\end{figure}

Figure~\ref{profiles} shows the radial profiles\footnote{ We derived the
  radial profiles using the IRAF task {\sl radprof}. This task fits a 
  point spread function on each star and 
    selected objects in the field through an input coordinate file, deriving
    a Full Width at Half Maximum of the fitted profile.}
of DRP-A539a and DRP-A634a, derived from our sharpest images.
The first is marginally different from the typical stellar profile,
showing a radius at half maximum $r_{FWHM} \simeq 0.44''$ ($\left<
  r_{FWHM}^{star}\right> \simeq 0.39''$). 
The second is almost indistinguishable from the stellar profile,
with $r_{FWHM} \simeq 0.39''$ ($\left< r_{FWHM}^{star}\right> \simeq 0.38''$). 
 After correcting these profiles for the effect of seeing under the 
  assumption of gaussian PSF\footnote{ \citealt{dri05} have reported
      a slight deviation of a gaussian behaviour when correcting
      effective radii of galaxies for the effect of seeing in the
      Millennium Galaxy Catalogue (MGC). For the scope of this work
      gaussianity behaviour of the PSF is assumed.}, 
  effective (half light) radii
  of 0.14'' and 0.09'' are derived, which correspond to 84 and 32~pc for DRP-A539a and 
  DRP-A634a respectively.
 These effective radii are lower than most values presented for 
  misclassified compact galaxies in the Millennium Galaxy Catalogue
  \citep{lis06} and resemble the values typically shown by UCDGs in
  nearby clusters \citep{drnk04}.  

 As concerns LSB-A634a, the length of its major axis
  from our continuum image was estimated to be $\approx 28''$. At 
  the distance of DRP-A634a this length corresponds to 16~kpc. This value is
  in  
good agreement with the dimensions of edge-on disc galaxies.
The angular size
  of the system argues against it being a high redshift object. If, for example,
  the emission line reported in Section~2.2 was one of the [O{\sc iii}]
  lines, the redshift would be $z \approx 0.35$, and the corresponding
  diameter of the system  at such a distance would be 138~kpc, which is
  highly unlikely for an edge-on galaxy. For the same reason, we can discard
  the possibility of 
  this system being located at larger distances.

 Broad band magnitudes can be derived for DRP-A634a since SDSS frames
  are available. Aperture photometry corrected for Galactic extinction of
  DRP-A634a yields   
  $M_{g'} = -15.81$~mag and $M_{r'} = -14.50$~mag. The
  broad band magnitudes of the composite system LSB-A634a$+$DRP-A634a were also derived,
  obtaining  $M_{g'} = -17.38$~mag and $M_{r'} = -17.30$~mag. Since
  DRP-A539a was observed with the {\sl Gunn r'} filter, 
a value of $M_{r'} \approx -16.07$~mag was estimated for this galaxy. 
By applying the average $g' - r' =
  -0.02$~mag  of the sample of Ultracompact Blue Dwarf Galaxies of  
  \citet{crb06}, a value of $M_{g'} \approx -16.09$ is obtained for DRP-A539a.
  The magnitudes derived for DRP-A539a and DRP-A634a are
  brighter than those typical of early-type UCDGs in clusters
  \citep{mie06}. Nevertheless, the Starburst99 model \citep{lei99} predicts
  that the optical magnitudes of an instantaneous burst of star formation can
  fade by more than 3 magnitudes in about 10$^8$ yr. These results together
  with the limits to the sizes of DRP-A539a and DRP-A634a, open the
  possibility of them being the progenitors of early-type UCDGs
  recently found in nearby clusters.

The equivalent widths of the most conspicuous Balmer emission lines of our
two objects are relatively high: EW(H$\beta$) (77 and 280\AA\ for DRP-A539a and
DRP-A634a respectively) are in fact above the median value ($\sim 40$\AA)
reported for the sample of H{\sc ii} galaxies of \citet{ter91}
and also for the sample of H{\sc i}-rich dwarf galaxies in the Hydra
cluster of \citet{duc01}. In addition, EW(H$\alpha + $[N{\sc ii}])
(510 and 1290 \AA\ respectively) are several times larger than the 
corresponding values of star
forming dwarf galaxies reported in other clusters such as Fornax 
\citep{drnk01}, Virgo \citep{bos02a,bos02b}, Coma \citep{ige02} and A1367
\citep{ige02,cor06}. 
 When comparing with the UCBDs of \citet{crb06} we find that the
  equivalent width of DRP-A539a is among the typical values of this 
  sample, but DRP-A634a remains an extreme case.
 The high equivalent widths shown by our two galaxies reveal very 
strong and young star formation activity. The ages and stellar
mass of the recent star formation episodes can be estimated from the
luminosities and 
equivalent widths of the Balmer lines. By assuming an instantaneous burst of
star formation, and using Starburst99 \citep{lei99} for a Kroupa IMF with
$M_{up} = 100$~M$_{\odot}$ and $M_{low} = 0.5,0.1$~M$_{\odot}$ and the Padova
AGB tracks, we find that the ages predicted are 5.8 and 4.5~Myr (5 and 4 Myr
using Geneva High tracks) for DRP-A539a and DRP-A634a respectively.  The
corresponding masses which are being converted into stars are 2.0 and
2.2~10$^6$M$_{\odot}$ respectively,  with an uncertainty of 15\%\footnote{The
  derived values for the age and stellar mass of the burst are the nominal
  values obtained from the measured imaging H$\alpha$ flux and corresponding 
  error (derreddening error is included).}. This computation assumes the
approximation that all Lyman continuum photons are absorbed by the gas. 

\begin{figure}
\centering
\includegraphics[width=14cm]{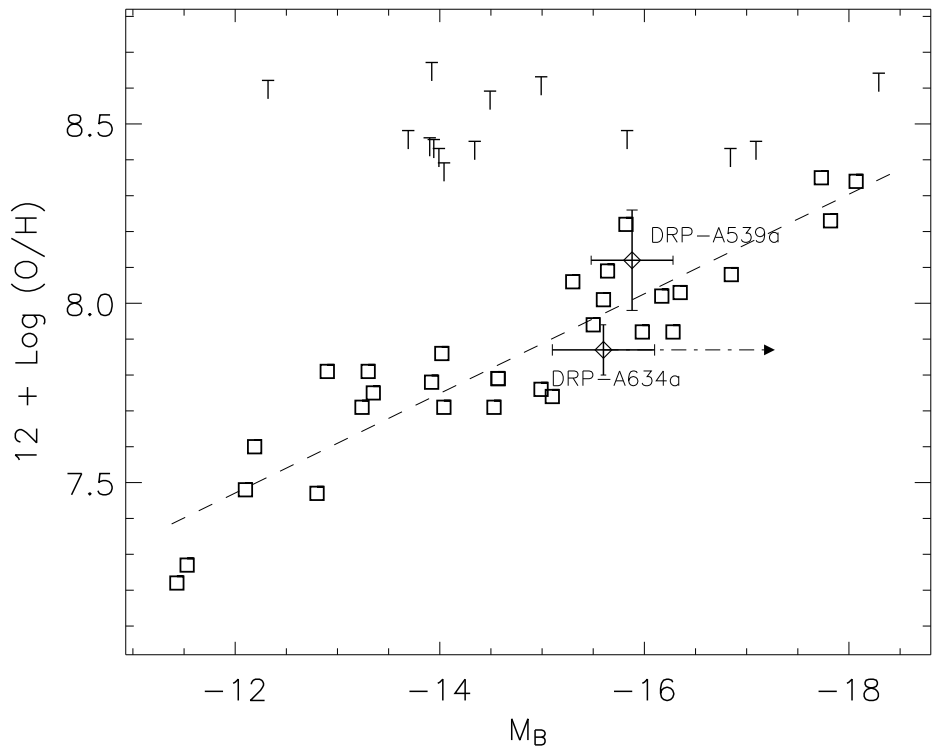}
\caption{Luminosity-Metallicity relation for nearby dwarfs galaxies (open
    squares) adapted from \citet{pil04}. ``T'' symbols  represent Tidal dwarfs
    (e.g. \citealt{kun00} and references therein). Our objects are
    overplotted as open  
    diamonds with their error bars. The position of the luminosity $M_{B}$ of
    the ensemble system LSB-A634a$+$DRP-A634a, is pointed out by the arrow.}
\label{MB_Z}
\end{figure}

 Figure~\ref{MB_Z} shows the luminosity-metallicity relation for
  dwarf galaxies (after \citealt{pil04}), including
  the points for DRP-A539a and DRP-A634a, with $M_{B} = -15.60$ mag and
  $-15.89$ mag respectively. $B$ absolute magnitudes have been obtained from
  $M_{g'}$ after  applying the correction term $g' - B = -0.21$, as reported
  by \citet{fuk95} for late-type dwarf galaxies. The two points follow the
  mean  relation reported for nearby dwarf galaxies and remain far from the
  locus  occupied by typical tidal dwarf galaxies (TDG) (see review by
  \citealt{kun00} and references therein).  In figure~\ref{MB_Z}, the point
  corresponding to the integrated system LSB-A634a$+$DRP-A634a appears to
  separate from the luminosity-metallicity relation by more than 1 mag,
  being too luminous for the metallicity derived for DRP-A634a (table
  \ref{spec_pro}).

\section{Discussion}

\subsection{DRP-A539a}

\begin{figure}
\centering
\includegraphics[width=14cm]{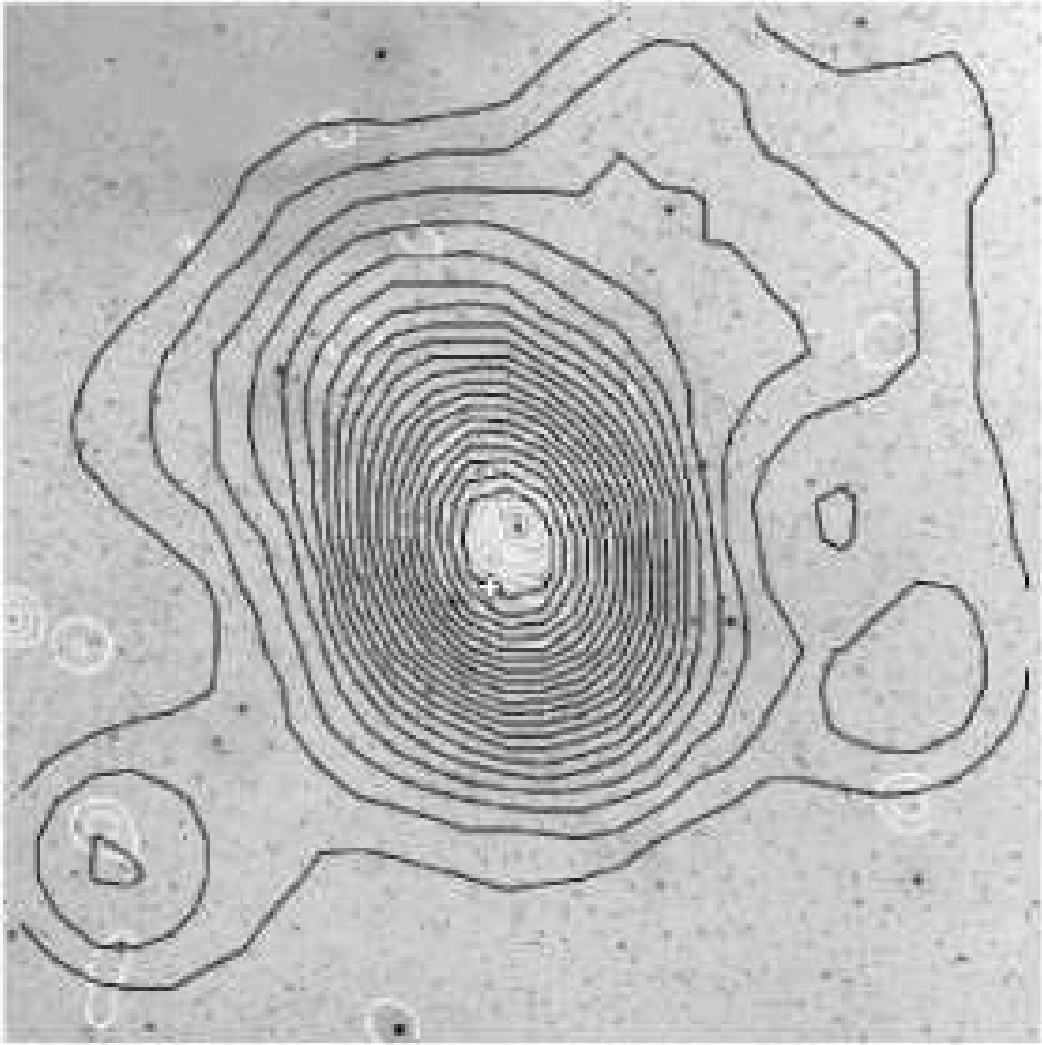}
\caption{Abell 539 cluster DSS image with X-Ray ROSAT detection (white) and
2MASS galaxy spatial-density contours (black) overplotted (2x2 degrees)
(relative units have been used for both contours parameters).
Cross marks the position of the compact starburst galaxy relative to the
cluster center. (North is to the top and East is to the left.)}
\label{A539encuad}
\end{figure}

Figure~\ref{A539encuad} shows the optical DSS (Palomar Digitalized Sky
Survey) frame of the inner region of the cluster A539. Contours corresponding
to the X-ray emission\footnote{From 0.4-2.4 Kev ROSAT map.} and to the
surface density of galaxies (from 2MASS) of A539 are overlaid. It can be seen
that the maxima of surface density of galaxies and X-ray are coincident. The
X-ray luminosity of A539 is 6.7 $10^{44}$ erg~s$^{-1}$ \citep{whi97}.
Its velocity dispersion was estimated to be 629~km~s$^{-1}$, thus giving a
total dynamical mass for the cluster of $32 \times 10^{13}$~M$_{\odot}$
(Struble \& Rood 1999). The projected distance of DRP-A539a from the  center
of A539 is 300~kpc, which corresponds to $0.2 \times r_{200}$. This short
distance together with the measured radial velocity of DRP-A539a, ensures
that this object is well within the region defined by the caustic of the
cluster  (see Figure~2 of \citealt{rin03}\footnote{In the
    particular case of A539, the caustic appears well defined and any
    possible contamination is low. However, these authors remark
    that although the caustic separates cluster members from foreground and
    background galaxies in a more efficient way than the velocity sigma
    clipping, still some interloppers may lie within the caustics.}).  

The existence of such an active star forming object at such a short distance
from the center of a cluster is unusual and opens several questions about
its nature and origin. One possibility is that this object could be a tidal
dwarf galaxy, resulting from a galaxy-galaxy interaction.
These kinds of objects have been previously reported in different
environments such as clusters \citep{duc99,duc00,ige03}, groups
\citep{ige01,men04} or just pairs of interacting galaxies \citep{mir92,duc00}.
TDGs originate out of the tidal features resulting during the
interaction, and under certain conditions they can escape the potential well
of the parent galaxy and evolve independently \citep{elm93}.
A visual inspection of the ON-band and OFF-band
frames shows no signatures of galaxy-galaxy interactions
around DRP-A539a. The closest galaxy is 2MASX J05165377+0619216, an S0 galaxy
whose radial velocity is $v = 8063$~km~s$^{-1}$, located 
at a projected distance of 32~arcsec (about 18~kpc).
We analyzed the isophotes of this galaxy and did not find any distortion or
abnormal twist of the isophotes that could be taken as a signature of
interaction.
Moreover, no diffuse emission was detected around this galaxy
in the direction of DRP-A539a. In addition, as shown in
  Figure~\ref{MB_Z}, DRP-A539a follows the mean
  luminosity-metallicity relation derived for a large sample of dwarf
  galaxies. Based on these arguments, we can discard DRP-A539a as being a TDG. 

Another possibility is that DRP-A539a is the product of
ICM-induced efficient star formation in gas clouds drifting
into the cluster \citep{bek03}. In this scenario
the star formation activity results from the compression of
molecular clouds stripped from spiral galaxies through
galaxy-galaxy or galaxy-tidal field interactions. Isolated
intra-cluster H{\sc ii} regions have already been reported
in the Virgo cluster \citep{ger02,cor04}.
However, these H{\sc ii} regions are more than one order of magnitude less
luminous than DRP-A539a. It should be noted that the objects studied in
\citet{cor04} are located close to the two bright spirals VCC~836 and 
VCC~873, which are probably their progenitors. Nonetheless, DRP-A539a does
not appear to be associated to any gas-rich bright galaxy, suggesting that either
the parent molecular cloud is the result of an ancient episode of gas
stripping, where the stripped galaxy is already far away, or rather we are
facing a star forming dwarf galaxy (SFDG) falling for the first time into the
cluster core.  Theoretical models predict that SFDGs lose their external
gas very rapidly when they  enter the cluster potential well, due to the ram
pressure exerted by the ICM \citep{mor00} and also, according to 
\citet{bek03},the timescale for transformation of gas into stars due to ICM
pressure is of the order of 10~Myr; for this reason we argue that DRP-A539a
is in the very early stages of the infall process.

\subsection{DRP-A634a}

\begin{figure}
\centering
\includegraphics[width=14cm]{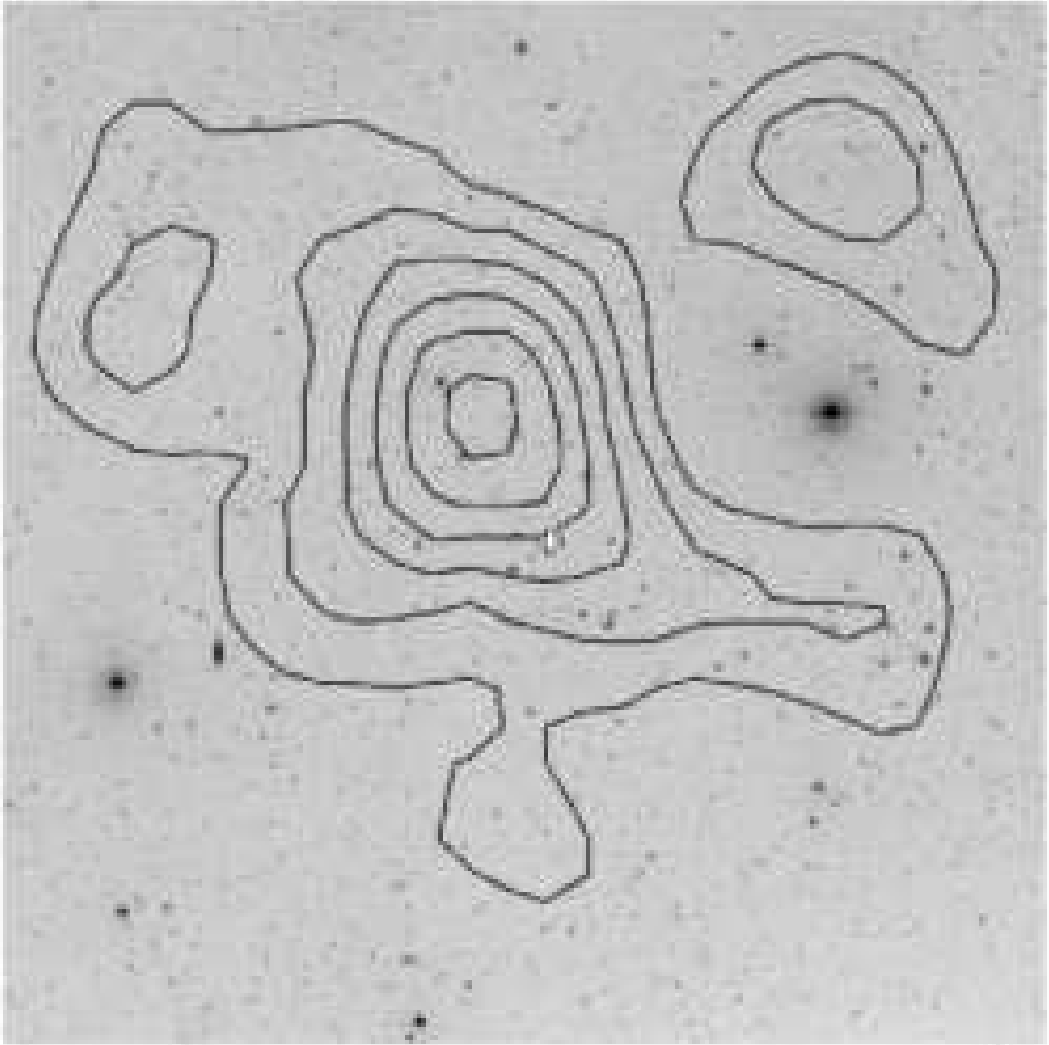}
\caption{Abell 634 overview (2x2 degrees) with the 2MASS galaxy
spatial-density contours (relative units, black). Cross marks the position of the compact
dwarf galaxy detected (North is to the top and East is to the
left).  Because of the extreme low X Ray emission detected by ROSAT for this
cluster, no X-Ray contours can be overplotted.} 
\label{A634encuad}
\end{figure}

Figure~\ref{A634encuad} shows the optical DSS frame of the cluster
A634 with the overlaid contours corresponding to the surface density of
galaxies derived from 2MASS. This cluster is not detected in X rays by the
ROSAT mission, so an upper limit of 6$\cdot$10$^{41}$ erg s$^{-1}$ is adopted
for its X-ray luminosity.  
The  non detection of X-ray emission means
that this cluster is probably in the process of formation and therefore a
virialized core is not yet in place. Nevertheless, the smoothed
 surface  density contours from 2MASS show a dense central aggregate
of galaxies showing the 
galaxy cluster location and shape.
  We must note that the cluster center as indicated by the
  2MASS-contours, $\alpha = 08$:$15$:$08$, $\delta = +58$:$14$:$58$,  
  is about 12 arcmin away from the center quoted in \citet{str99}. Hereafter,
  for our dynamical considerations and being conservative, we have
  adopted the center of  the cluster inferred by the 2MASS contours. The
  heliocentric velocity 
  measured  for DRP-A634a differs from the cluster heliocentric velocity by
  $1.34 \times \sigma_{A634}$ km~s$^{-1}$, and this object is located at a
  projected  distance of 398~kpc from the cluster center.
Both  values, heliocentric velocity and projected distance, would place
DRP-A634a  within the caustic of every galaxy cluster presented in
\citet{rin03}, 
at the $2~\sigma$ confidence level\footnote{ In the case the cluster center
  quoted by \citet{str99} is adopted, the corresponding projected  distance
  would amount to 16 kpc and  DRP-A634a will be a cluster member at a higher
  confidence level.}; Especially in the case of A194, a cluster presenting
velocity dispersion and extension in the sky very similar to those of
A634. The non detection in X-ray of A634 together with the substantial
distance of DRP-A634a to the cluster center, may weaken the hypothesis that
the strong star formation activity shown by this object could be the result of
compression of molecular clouds due to the pressure exerted by the
ICM. Nonetheless, we 
must bear in mind that the lack of detection of X-Ray emission in A634
does not necessarily mean it is devoid of a significant ICM.

 The comparable radial velocities measured for DRP-A634a and LSB-A634a,
  and their close position in the sky could be indicative of both objects
  being physically related. The optical morphology of the
  LSB-A634a$+$DRP-A634a system may suggest an edge-on, late-type,
low surface brightness galaxy, with
  an off-center very bright knot accounting for
  most of the optical light. In this scenario, would this system be
  anywhere close to the Tully-Fisher (TF) relation? To answer this question,
  first we have assumed that the difference between the radial velocities of DRP-A634a and 
  LSB-A634a{\_}knot provides an estimation of the internal velocity of the system,
  giving $2~V_{max} \approx 80$  km~s$^{-1}$. Secondly, we have applied the
  results of \citet{pie99}, who derived the TF relation for a sample including dwarf 
  star-forming galaxies from the Virgo Cluster Catalogue in the $K'$ band. 
  The $M_{K'}$ magnitude of the system was derived from the $M_{B}$ magnitude
  (see sect. 3) making use of: 1) the average $(B - K)$ colour for the Virgo
  sample of 57 ({\sl Sdm-Sd/Sm} to {\sl Im/BCD}) galaxies\footnote{
 The GOLDmine database \citep{gav03b} was used to derive these data -
    http://goldmine.mib.infn.it/ }:  $(B - K) \simeq  2.85 $;
  and 2) the mean colour, $(B - K_{s}) \simeq 2.20$, obtained for the sample
  of BCDs from \citet{noe03}. Taking the average value of both results, $<(B
  - K)> = 2.53 \pm 0.33$,   we found  $M_{K'} = -19.70 \pm 0.35$~mag for
  the system LSB-A634a$+$DRP-A634a. Applying  the fit of \citet{pie99} (see
  their figure~6) to our value of $V_{max}$, their TF relation   
  predicts $M_{K'} = -16.89$~mag, nearly 3~mag fainter than the value of
  $M_{K'}$  estimated above for the whole system. 
  In the optical, several papers on the TF relation for low luminosity
  galaxies have appeared recently.
  The TF relation for local discs and
  irregular galaxies from \citet{zie02} would predict a value of $M_{B} \sim
  -15$ mag 
  for our $V_{max} \approx 40$ km~s$^{-1}$, about 2 mag fainter than the
  value measured for our system.
  According to \citet{swa99} (figure 13, page 117), galaxies with absolute
  magnitudes fainter than $M_{r'} \simeq -18$ systematically fall below a
  straight line TF relation.  
  This fact has been highlighted by \citet{mcg00} pointing out the different 
  evolutionary stage of  gas-rich dwarfs with respect to spiral discs. More 
  recently \citet{sch06} concluded that dwarf galaxies form a distinct
  sequence, being more diffuse than disc galaxies. 
  To summarize, the predictions of the TF relation for galaxies with internal
  velocity of order $V_{max} \approx 40$ km~s$^{-1}$ yield
  luminosity values much fainter than that determined here for this system. 
  On the other side, as mentioned in sect. 3 this integrated system separates
  from the luminosity-metallicity relation being too luminous for its
  metallicity.  According to these findings, the object formed by
  LSB-A634a$+$DRP-A634a is not proven to be a single galaxy.

 Even though the LSB-A634a$+$DRP-A634a system is not likely to be a single
  galaxy, still both objects could be physically related. Under this 
  assumption, we propose that the strong star formation activity shown by
  DRP-A634a is the result of the encounter between these objects. In
  fact, examples of recent star formation bursts associated to small groups
of galaxies which are falling into a cluster have already been
reported in the literature \citep{sak02,gav03,cor06},
and are thought to be associated to the so-called preprocessing of galaxies
before entering the cluster environment. This mechanism could account
for a non negligible fraction of the evolution of galaxies in dense
environments.  Further observations of this intriguing system are needed
  in order to  fully understand its nature and evolutionary
  state.
 
\section{Conclusions}

We have reported two examples of extreme star forming objects in
nearby clusters with different properties and at different
evolutionary phases. The observations show that albeit they
do not share the same origin, both compact and young starbursts show
a very intense star formation activity, even when compared to similar objects
in other nearby clusters or in less dense environments.

The origin of the two starbursts reported in this paper are
probably associated to different physical mechanisms: DRP-A539a
is directly associated to the dense and hot ICM which compresses
intergalactic clouds and induces star formation episodes in a short
timescale, before ram pressure is able to sweep the external gas. The
case of DRP-A634a can be related to the so-called ``preprocessing'' of
galaxies before they enter the cluster environment. In this case, the
aggregates of galaxies whose final fate is to fall into the cluster
inner regions, are the environments where secondary evolution is
taking place.

Two questions arise from these considerations: what is the relative
importance of such compact and extreme starbursts with respect to the global
SFR of nearby clusters? and, is there any evolutionary link between them and
the early-type ultracompact dwarf galaxies already reported in the
literature? \citep{drnk04}
To answer the first question, a detailed search based on H$\alpha$ surveys is
required. We note that an extensive spectroscopic survey of Abell 539 devoted
to studying the star formation in cluster galaxies has been carried out by
\citet{rin05}, but DRP-A539a was not selected there
because their sample of galaxies was NIR-magnitude limited. 
The same would have happened with DRP-A634a if the cluster A634 had
been surveyed under the same conditions. These non detections are naturally
explained by the fact that these objects are dwarfs and very young. However,
they show up very easily in wide field H$\alpha$ imaging surveys despite
their size and small stellar content. These kinds of surveys are required to
carry out a detailed census of compact starbursts in clusters of galaxies. In
this way, their relative contribution to the total SFR budget of nearby
clusters will definitely be determined.
As concerns the second question, several explanations have already been
proposed for the origin of UCDGs (see \citealp{jon06} for an interesting
review), although the discussion remains open. We propose that compact and
strong starbursts like the ones presented in this paper could evolve to
early-type dwarf galaxies after cessation of star formation and, if stripping
is efficient as the galaxy approaches the innermost regions of the cluster,
become UCDGs like the ones reported in Virgo and Fornax clusters.
A complete census of compact starbursts in clusters, with accurate projected
positions and surface densities will also help to answer this question.

\acknowledgments
 We thank the anonymous referee for his/her useful comments and
  suggestions. We also want to thank Steve Donegan for his careful revision
  of the english expression in this paper.
The data presented in this paper were obtained using ALFOSC, which is owned
by the Instituto de Astrof\'\i sica de Andaluc\'\i a (IAA) and operated at
the Nordic Optical Telescope under agreement between IAA and the NBIfAFG of
the Astronomical Observatory of Copenhagen.
 This article is based on observations made with the Isaac Newton
  Telescope (INT) operated on the island of La Palma (Canary Island) by the
  Isaac Newton Group (ING) in the spanish Observatorio del Roque de los
  Muchachos. This research has made use of the NASA/IPAC Extragalactic
Database (NED) which is operated by the Jet Propulsion Laboratory, California
Institute of Technology, under contract with the National Aeronautics and
Space Administration.  This research has made use of the GOLDMine
  Database \citep{gav03b}. This publication makes use of data products from
the Two Micron All Sky Survey, which is a joint project of the University of
Massachusetts and the Infrared Processing and Analysis Center/California
Institute of Technology, funded by the National Aeronautics and Space
Administration and the National Science Foundation. This work has been
partially funded by the projects AYA2004-08260-C03-02, of spanish PNAYA, and
TIC114 of the Junta de Andaluc\'\i a. 

\clearpage

\appendix

\section{Appendix material}

\begin{center}
\begin{deluxetable}{llcc}
\tablewidth{10cm}
\tablecaption{Imaging observations log.}
\tablehead{\colhead{Object} & \colhead{Filter} & \colhead{Exp. time} &
  \colhead{Date}\\
\colhead{} & \colhead{} & \colhead{(s)} & \colhead{} \\
}
\startdata
DRP-A539a & ON           & $5 \times 600$   &  3 Dec. 2002  \\
          & ON           & $1 \times 1200$  &  3 Dec. 2002  \\
          & ON           &  $ 1200$         &  4 Dec. 2002  \\
          & OFF          & $2 \times 1200 $ &  28 Feb. 2003 \\
          & OFF          & $5 \times 600 $  &  28 Feb. 2003 \\
          & OFF          & $ 1200 $         &  4 Dec. 2002  \\
          &  r'          & $3 \times 300$   &  3 Dec. 2002  \\
\tableline
DRP-A634a & ON & $ 1200 $         &  9 Feb. 1999  \\
          & ON & $2 \times 1200$  &  26 Feb. 2003 \\
          & ON & $3 \times 400$   &  28 Feb. 2003 \\
          & OFF &  $ 1200 $        &  9 Feb. 1999  \\
          & OFF & $2 \times 1200$  &  26 Feb. 2003 \\
          & OFF & $4 \times 400$   &  28 Feb. 2003 \\
\enddata
\label{image_log}
\end{deluxetable}
\end{center}

\clearpage

\begin{center}
\begin{deluxetable}{lcccrr}
\tablewidth{17.3 cm}
\tablecaption{Basic photometric properties of DRP-A539a and DRP-A634a measured from our H$\alpha$ and continuum frames: (1)
  Object Id; (2) R.A.; (3) Dec.; (4) H$\alpha +$[N{\sc ii}] luminosity corrected for Galactic extinction; 
(5) H$\alpha +$[N{\sc ii}] equivalent width}
\tablehead{\colhead{Object} & \colhead{R.A.} & \colhead{Dec.} &
\colhead{$L$(H$\alpha +$[N{\sc ii}])} & \colhead{EW(H$\alpha +$[N{\sc ii}])} \\
 \colhead{} & \colhead{{\scriptsize (J2000)}} & \colhead{{\scriptsize (J2000)}} &
\colhead{{\scriptsize (10$^{40}$ erg~s$^{-1}$)}} & \colhead{{\scriptsize (\AA)}} }
\startdata
DRP-A539a             &  $05$:$16$:$51.7$ &  $+06$:$19$:$32.1$ & 1.27 $\pm$ 0.03 & { }430 $\pm$ { }60 \\
DRP-A634a             &  $08$:$13$:$55.6$ &  $+58$:$02$:$32.4$ & 2.38 $\pm$ 0.21 & 1010 $\pm$ 230 \\
LSB-A634a{\_}knot     &  $08$:$13$:$57.0$ &  $+58$:$02$:$42.5$ & 1.03 $\pm$ 0.25 & { }{ }65 $\pm$ { }23 \\
LSB-A634a$+$DRP-A634a &                   &                    & 3.4{ } $\pm$ 0.4{ } & { }190 $\pm$ { }40 \\
\enddata
\label{phot_pro}
\end{deluxetable}
\end{center}

\clearpage

\begin{deluxetable}{lccccc}
\tablewidth{18cm}
\tablecaption{Spectroscopy Observations log.}
\tablehead{\colhead{Object} & \colhead{Date} & \colhead{Spect. Range} &
  \colhead{Exp. Time} & \colhead{Slit width} & \colhead{Position angle$^{\dagger}$}\\
\colhead{} & \colhead{} & \colhead{(\AA)} & \colhead{(s)} & \colhead{(arcsec)} & \colhead{(deg)}\\
}
\startdata
DRP-A634a & 2004-11-19 & 3700-6100 & $1 \times 1800$ & 1.2   & 245.5 \\
DRP-A634a & 2004-11-19 & 6000-8000 & $1 \times 1800$ & 1.2   & 235.5 \\
DRP-A539a & 2004-12-09 & 3700-6100 & $3 \times 1800$ & 0.4   & 336.4, 356.4,
15.4 \\
DRP-A539a & 2004-12-09 & 6000-8000 & $3 \times 1800$ & 0.4   & 312.7, 323.1,
30.4 \\ 
{\footnotesize LSB-A634a$+$DRP-A634a} & 2004-12-09 & 6000-8000 & $1 \times 600 $ & 0.4   & 54.0 \\

\enddata

\tablecomments{ $^{\dagger}$Position angle (measured from North to
  East) corresponds for every value on the table to the Parallactic Angle
  except the last exposure.} 
\label{spectra_log}
\end{deluxetable}

\clearpage

{\scriptsize
\begin{deluxetable}{cccrr}
\tiny
\tablewidth{14.5cm}
\tablecaption{Reddening corrected  line intensities of the Objects
DRP-A539a and DRP-A634a
relative to H$\beta$=1000. The reddening coefficient C(H$\beta$), electron
temperatures, oxygen abundances,  as well as the H$\beta$ flux and the
equivalent width and for H$\beta$, H$\alpha$ and [O{\sc ii}] are
quoted. H$\beta$ flux has been corrected for Galactic and intrinsic  
extinction using the extinction law $R = 3.1$ and their C(H$\beta$) }
\tablehead{
\colhead{Line} & \colhead{$\lambda$} & \colhead{f$_\lambda$} &
\colhead{DRP-A539a} &  \colhead{DRP-A634a}  \\
\colhead{} & \colhead{({\AA})} &\colhead{} & \colhead{} &
\colhead{} }
\startdata
  $[$O{\sc ii}$]$         & 3727  &  0.28 & 1058 $\pm$ { }7  &   375  $\pm$  17  \\
  $[$Ne{\sc iii}$]$       & 3868  &  0.24 &  522 $\pm$ 14    &   419  $\pm$  80  \\
  H{\sc i}                & 3889  &  0.24 &  210 $\pm$ 19    &   112  $\pm$  15  \\
  H{\sc i}                & 3970  &  0.22 &  320 $\pm$ 30    &   258  $\pm$  17  \\
  $[$S{\sc ii}$]$         & 4068  &  0.20 &   77 $\pm$ 15    &         ------         \\
  H$\delta$               & 4101  &  0.19 &  268 $\pm$ 15    &   171  $\pm$  22  \\
  H$\gamma$               & 4340  &  0.13 &  513 $\pm$ 10    &   473  $\pm$  21  \\
  $[$O{\sc iii}$]$        & 4363  &  0.13 &   70 $\pm$ 20    &   202  $\pm$  16  \\
  He{\sc i}               & 4471  &  0.10 &       ------     &    53  $\pm$  13  \\
  $[$Ar{\sc iv}$]$        & 4711  &  0.04 &       ------     &    59  $\pm$  14  \\
  H$\beta$                & 4861  &  0.00 & 1000 $\pm$ 16    &  1000  $\pm$  10  \\
  $[$O{\sc iii}$]$        & 4959  & -0.04 & 1869 $\pm$ { }7  &  2740  $\pm$  14  \\
  $[$O{\sc iii}$]$        & 5007  & -0.05 & 5280 $\pm$ 40    &  8272  $\pm$  10  \\
  He{\sc i}               & 5876  & -0.26 &   73 $\pm$ { }8  &   105  $\pm$  10  \\
  H$\alpha$               & 6563  & -0.37 & 2827 $\pm$ { }5  &  2789  $\pm$  { }5  \\
  $[$N{\sc ii}$]$         & 6584  & -0.37 &   47 $\pm$ { }5  &    27  $\pm$  { }5  \\
  He{\sc i}               & 6678  & -0.38 &     ------       &    29  $\pm$  { }5  \\
  $[$S{\sc ii}$]$         & 6717  & -0.39 &  112 $\pm$ 10    &    46  $\pm$  { }5  \\
  $[$S{\sc ii}$]$         & 6731  & -0.39 &  130:            &    29  $\pm$  { }5  \\
   He{\sc i}              & 7065  & -0.43 &       ------     &    58  $\pm$  { }7  \\
  $[$Ar{\sc iii}$]$       & 7135  & -0.44 &       ------     &    85  $\pm$  { }9  \\ \tableline
  C(H$\beta$)             &       &       & 0.27             & 0.47 \\ 
T$_e$([O{\sc iii}])       & (K)   &       & 12800 $\pm$ 1500 & 16800 $\pm$     700   \\
T$_e$([O{\sc ii}])        & (K)   &       & 11900 $\pm$ 1000 & 14800  $\pm$    500   \\
12+log(O$^{+}$/H$^{+}$)   &       &       & 7.65 $\pm$ 0.14  & 6.75  $\pm$     0.06 \\
12+log(O$^{++}$/H$^{+}$)  &       &       & 7.93 $\pm$ 0.14  & 7.83  $\pm$     0.05 \\
12+log(O/H)               &       &       & 8.12 $\pm$ 0.14  & 7.87  $\pm$ 0.06 \\
F(H$\beta$)& {\footnotesize (10$^{-15}$ erg cm$^{-2}$ s$^{-1}$)}   & & 4.21 $\pm$ 0.09 & 8.75 $\pm$ 0.08 \\
EW(H$\beta$)& ({\AA})  & & 77 $\pm$ 17{    } & 280 $\pm$ { }90{  } \\
EW(H$\alpha$)& ({\AA})  & & 510 $\pm$ 90{    } & 1290 $\pm$ 120{ } \\
EW([O{\sc ii}])& ({\AA})  & & 70 $\pm$ 15{    } & 110 $\pm$ { }40{  } \\

\enddata                                                             

\label{spec_pro}
\end{deluxetable}}
\normalsize

\newpage

\clearpage

\clearpage

\clearpage

\clearpage

\clearpage


\begin{thebibliography}{}

\bibitem[Arnaboldi et al.(2003)]{arn03} Arnaboldi, M., Freeman, K.C.,
  Okamura, S., et al. 
2003, \aj, 125, 514

\bibitem[Balogh et al.(1998)]{bal98} Balogh, M. L., Schade, D., Morris,
  S. L., Yee, H. K. C., Carlberg, R. G., and Ellingson, E. 1998, \apj, 504, L75

\bibitem[Balogh et al.(2004)]{bal04}    Balogh, M. L., Baldry, Ivan K.,
  Nichol, R., Miller, C., Bower, R., and Glazebrook, K. 2004, \apj, 615, L101

\bibitem[Bekki \& Couch(2003)]{bek03} Bekki, K., and Couch, W. J.
2003, \apj, 596, L13


\bibitem[Boselli \& Gavazzi(2002)]{bos02a} Boselli, A., and Gavazzi, G. 2002,
A\&A,  386, 124

\bibitem[Boselli et al.(2002)]{bos02b} Boselli, A., Iglesias-P\'aramo, J.,
V\'\i lchez, J. M., and Gavazzi, G. 2002, A\&A,  386, 143

\bibitem[Boselli \& Gavazzi(2006)]{bos06} Boselli, A., and
Gavazzi, G.\ 2006, \pasp, 118, 517


 \bibitem[Cardelli et al.(1989)]{car89} Cardelli, J.~A., 
Clayton, G.~C., and Mathis, J.~S.\ 1989, \apj, 345, 245 

 \bibitem[Corbin et al.(2006)]{crb06}  Corbin, M. R.,
  Vacca, W. D., Cid Fernandes, R., Hibbard, J. E., Somerville, R. S.,
  Windhorst, R. A. accepted for publication in ApJ, astro-ph/0607280

\bibitem[Cortese et al.(2004)]{cor04} Cortese, L., Gavazzi, G., Boselli, A.,
and Iglesias-P\'aramo, J. 
2004, A\&A, 416, 119

\bibitem[Cortese et al.(2006)]{cor06} Cortese, L., Boselli, A., Buat, V.,
  Gavazzi, G., Boissier, S., Gil de Paz, A., Seibert, M., Madore, B. F., and
  Martin, D. C. 2006, \apj, 637, 242

\bibitem[Drinkwater \& Hardy(1991)]{drnk91} Drinkwater, M., and Hardy, E. 1991,
  \aj, 101, 94

 \bibitem[Drinkwater et al.(2000)]{drnk00} Drinkwater, M.~J., 
Jones, J.~B., Gregg, M.~D., and Phillipps, S.\ 2000, Publications of the 
Astronomical Society of Australia, 17, 227 

\bibitem[Drinkwater et al.(2001)]{drnk01} Drinkwater, M. J., Gregg, M. D.,
Holman, B. A., and Brown, M. J. I. 2001, {\mnras}~ 326, 1076

\bibitem[Drinkwater et al.(2004)]{drnk04} Drinkwater, M. J., Gregg, M. D.,
  Couch, W. J., Ferguson,  H. C., Hilker, M., Jones, J. B., Karick, A., and
  Phillipps, S. 2004, PASA, 21, 375

 \bibitem[Driver et al.(2005)]{dri05} Driver, S.~P., Liske, 
J., Cross, N.~J.~G., De Propris, R., and Allen, P.~D.\ 2005, \mnras, 360, 81 



\bibitem[Duc et al.(1999)]{duc99} Duc, P. A., Papaderos, P., Balkowski, C.,
Cayatte, V., Thuan, T. X., and  van Driel, W.
1999, A\&ASS, 136, 539

\bibitem[Duc et al.(2000)]{duc00} Duc, P. A., Brinks, E., Springel, V., Pichardo, B.,
Weilbacher, P., and  Mirabel, I. F.
2000, \aj, 120, 1238

\bibitem[Duc et al.(2001)]{duc01} Duc, P. A., Cayette, C., Balkowski, C.,
  Thuan, T. X., Papaderos, P., and  van Driel, W. 
2001, \aj, 369, 763

\bibitem[Elmegreen, Kaufman \& Thomasson(1993)]{elm93}  Elmegreen, B.G.,
  Kaufman, M., and Thomasson, M. 1993, \apj, 412, 90

\bibitem[Freeman et al.(2000)]{fre00}  Freeman, K. C., Arnaboldi, M., Capaccioli,
  M., Ciardullo, R., Feldmeier, J., Ford, H., Gerhard, O., Kudritzki, R.,
  Jacoby, G., M\'endez, R. H., and Sharples, R. 2000, ASPC, 197, 389

\bibitem[Fujita \& Nagashima(1999)]{fuj99} Fujita, Y., and Nagashima, M.
1999, \apj, 516, 619

 \bibitem[Fukugita et al.(1995)]{fuk95} Fukugita, M., 
Shimasaku, K., and Ichikawa, T.\ 1995, \pasp, 107, 945 

\bibitem[Gallagher \& Hunter(1989)]{gal89} Gallagher, J. S., III, and  Hunter,
  D. A. 1989, \aj, 98, 806G

\bibitem[Gavazzi et al.(2002)]{gav02} Gavazzi, G., Boselli, A., Pedotti, P.,
  Gallazzi, A., and Carrasco, L.  2002, A\&A, 396, 449

\bibitem[Gavazzi et al.(2003a)]{gav03} Gavazzi, G., Cortese, L., Boselli, A.,
  Iglesias-P\'aramo, J., V\'\i lchez, J. M., and  Carrasco, L. 2003a, \apj,
  597, 210 

 \bibitem[Gavazzi et al.(2003b)]{gav03b} Gavazzi, G., Boselli, 
A., Donati, A., Franzetti, P., and Scodeggio, M.\ 2003b, \aap, 400, 451 

\bibitem[Gerhard et al.(2002)]{ger02} Gerhard, O., Arnaboldi, M., Freeman,
  K. C., and Okamura, S.
2001, \aj, 581, 109

\bibitem[Gerhard et al.(2005)]{ger05}   Gerhard, O., Arnaboldi, M.,
  Freeman, K. C., Kashikawa, N., Okamura, S., and Yasuda, N.
 2005, ApJ, 621, L93

\bibitem[Hummer \& Storey(1987)]{hum87} Hummer, D. G., and Storey, P. J. 1987
\mnras, 224, 801

\bibitem[Iglesias-P\'aramo \& V{\'\i}lchez(2001)]{ige01} Iglesias-P\'aramo,
  J., and  V\'\i lchez, J. M. 
2001, \apj, 550, 204

\bibitem[Iglesias-P\'aramo et al.(2002)]{ige02}  Iglesias-P\'aramo, J.,
  Boselli, A., Cortese, L., V\'\i lchez, J. M., and Gavazzi, G.
2002, A\&A, 384, 383

\bibitem[Iglesias-P\'aramo et al.(2003)]{ige03} Iglesias-P\'aramo, J., van
  Driel, W., Duc, P. A., Papaderos, P., and V\'\i  lchez, J. M. et al
2003, A\&A, 406, 453

\bibitem[Jones et al.(2006)]{jon06} Jones, J. B, Drinkwater, M. J., Jurek, R.,
  Phillips, S., Gregg, M. D., Bekki, K., Couch, W. J., et al. 2006, \aj, 131, 312

\bibitem[Kunth \& \"{O}stlin(2000)]{kun00} Kunth, D., and \"{O}stlin,
  G. 2000, A\&ARv, 10, 1

\bibitem[Lee et al.(2003)]{lee03} Lee, H., McCall, M.~L., and
Richer, M.~G.\ 2003, \aj, 125, 2975

\bibitem[Leitherer et al.(1999)]{lei99} Leitherer, C., Shaerer, D., Goldader,
  J. D., Gonz\'alez-Delgado, R. M., et al 1999, AJSS, 123, 3

\bibitem[Lewis et al.(2002)]{lew02} Lewis, I., Balogh, M., De Propris,
  R., Couch, W., Bower, R., Offer, A., Bland-Hawthorn,
  J., Baldry, I. K., et al. 2002, MNRAS, 334, 673

\bibitem[Liske et al.(2006)]{lis06} Liske, J., Driver, S. P., Allen, P. D.,
Cross, N. J. G., and De Propris, R., submitted to MNRAS, astro-ph/0604211

 \bibitem[Noeske et al.(2003)]{noe03} Noeske, K.~G., 
Papaderos, P., Cair{\'o}s, L.~M., and Fricke, K.~J.\ 2003, \aap, 410, 481 

\bibitem[McGaugh(2000)]{mcg00} McGaugh, S.\ 2000, Bulletin of 
the American Astronomical Society, 32, 1496 

\bibitem[Mendes de Oliveira et al.(2004)]{men04} Mendes de Oliveira, C.,
Cypriano, E. S., Sodr\'e Jr., L., and Balkowski, C.
2004, \apj 605, L17

\bibitem[Mieske et al.(2006)]{mie06} Mieske, S., Hilker, M., Infante, L., and
  Jord\'an, A.  2006, \aj, 131, 2442

\bibitem[Mirabel, Dottori \& Lutz (1992)] {mir92} Mirabel, I.F., Dottori,
  H., and Lutz, D. 1992, A\&A, 256, L19

\bibitem[Moore et al.(1999)]{moo99} Moore, B., Lake, G., Quinn, T., and Stadel,
  J. 
1999,  \mnras, 304, 465

\bibitem[Moore et al.(2000)]{moo00} Moore, B., Quilis, V., and Bower, R. 2000,
  ASPC, 197, 363

\bibitem[Mori et al.(2000)]{mor00} Mori, M., and Burkert, A.
2000, \apj, 538, 559


 \bibitem[Pierini \& Tuffs(1999)]{pie99} Pierini, D., and
Tuffs, R.~J.\ 1999, \aap, 343, 751 

\bibitem[Pilyugin, V\'\i lchez, \& Contini(2004)]{pil04} Piluygin, L. S.,
  V\'\i lchez, J. M., and  Contini, T. 2004, A\&A, 425, 849


\bibitem[Poggianti(2004)]{pog04} Poggianti, B. 2004,  
{\sl Baryons in Dark matter halos}. Novigrad, Croatia, 5-9 Oct
2004. Proceedings of Science, http://pos.sissa.it, p.104.1   

\bibitem[Rines et al.(2003)]{rin03} Rines, K., Geller, M., Kurtz, M., and
Diaferio, A. 
2003, \aj, 126, 2152


\bibitem[Rines et al.(2005)]{rin05} Rines, K., Geller, M., Kurtz, M., and
Diaferio, A. 
2005, \aj, 130, 1482

\bibitem[Sakai et al.(2002)]{sak02} Sakai, S., Kennicutt, R. C., van der
  Hulst, J. M. and Moss, C. 2002, \apj, 578, 842

\bibitem[Schlegel, Finkbeiner \& Davis(1998)]{sch98} Schlegel, D. J.,
  Finkbeiner, D. P., and Davis, M. 1998, \apj, 500, 525

 \bibitem[Schombert(2006)]{sch06} Schombert, J.~M.\ 2006, \aj, 
131, 296 

 \bibitem[Struble \& Rood(1999)]{str99} Struble, M.~F., and
Rood, H.~J.\ 1999, \apjs, 125, 35

\bibitem[Swaters(1999)]{swa99} Swaters, R.~A.\ 1999, 
Ph.D.~Thesis. Gr\"oningen University.



\bibitem[Terlevich et al.(1991)]{ter91} Terlevich, R.,
Melnick, J., Masegosa, J., Moles, M., and Copetti, M.~V.~F.\ 1991, \aaps,
91, 285

\bibitem[Thuan \& Martin(1981)]{thu81} Thuan, T.~X., and
Martin, G.~E.\ 1981, \apj, 247, 823

\bibitem[Verdes-Montenegro et al.(2002)]{ver02} Verdes-Montenegro, L.,
Del Olmo, A., Iglesias-P\'aramo, J., Perea, J., V{\'\i}lchez, J. M.,
Yun, M. S., and Hutchmeier, W. K. 
2002, A\&A, 396, 815

\bibitem[V\'\i lchez(1995)]{vil95} V\'\i lchez, J. M. 1995, \aj, 110, 1090

\bibitem[V\'\i lchez \& Iglesias-P\'aramo(2003)]{vil03} V\'\i lchez,
  J. M., and Iglesias-P\'aramo, J. 2003, \apjs, 145, 225

\bibitem[White, Jones \& Forman(1997)]{whi97} White, D. A., Jones, C., and Forman, W.
1997, MNRAS, 292, 419

 \bibitem[Ziegler et al.(2002)]{zie02} Ziegler, B.~L., et al.\ 
2002, \apjl, 564, L69 

\end{thebibliography}
\end{document}